# Observing solar vortices with existing and future instrumentation.
# Solar Physics International Network for Swirls (SPINS) white paper (Helio)

**Authors and contact details**

| | | |
|---|---|---|
| Suzana Silva | suzana.silva@sheffield.ac.uk | University of Sheffield |
| Viktor Fedun | v.fedun@sheffield.ac.uk | University of Sheffield |
| Gary Verth | g.verth@sheffield.ac.uk | University of Sheffield |
| Istvan Ballai | i.ballai@sheffield.ac.uk | University of Sheffield |
| Eamon Scullion | eamon.scullion@northumbria.ac.uk | Northumbria University |
| Malcolm Druett | m.druett@sheffield.ac.uk | University of Sheffield |
| Kostas Tziotziou | kostas@noa.gr | National Observatory of Athens, Greece |
| Alex Pietrow | apietrow@aip.de | Leibniz-Institut für Astrophysik Potsdam, Germany |

A full list of co-authors / signatories is appended at the end

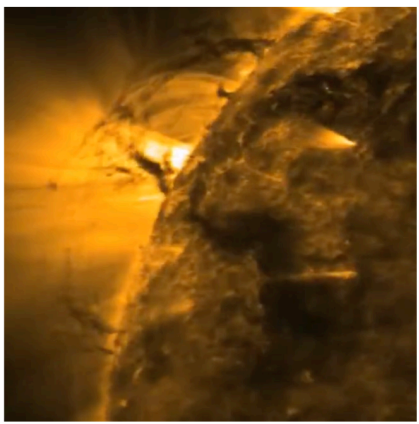
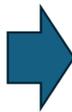
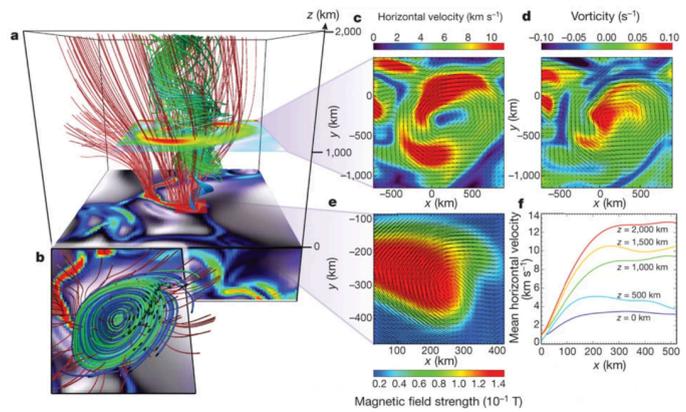
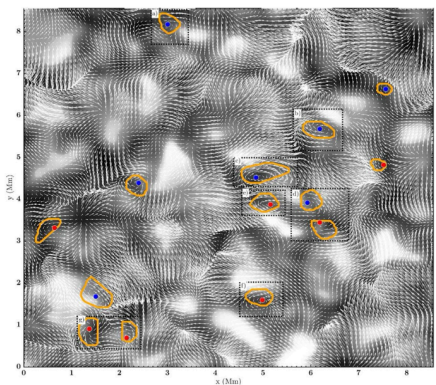
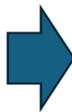
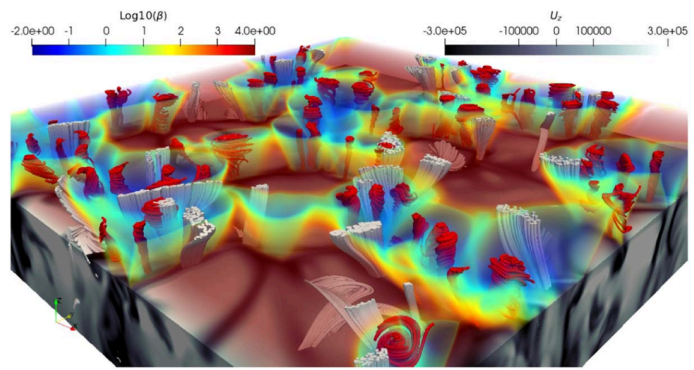

*Images courtesy: SDO, SST, Silva et al., 2021, Giagkiozis et al., 2018, Wedemeyer-Böhm S. et al., 2012.*

# 1. Scientific Motivation and Objectives

Solar vortices serve as natural environments for magnetic field twisting [1-10], energy concentration [11,12] and transport [11,13-22], flow coupling across atmospheric layers, and MHD wave guidance. State-of-the-art MHD simulations identify vortices as key drivers of energy transport (e.g., [10]). Numerical modelling and limited observational studies also increasingly relate them to solar atmospheric heating [15-22], jet formation triggering [23-28], solar wind acceleration [29], and small-scale energy release events [20, 21, 30, 31]. Despite their significance, our present understanding is fragmented, and our current knowledge is shaped by some critical limitations:

- *Observational challenges:* Solar vortices remain extremely difficult to detect and track reliably because their observable signatures depend sensitively on wavelength, formation height, local magnetic and radiation environment, temporal, and spatial resolution. Many fade or disappear entirely when the observing diagnostics change. No observational framework yet exists that can track vortices across atmospheric layers or disentangle vortical flows from similar flow phenomena. Current observations do not provide high-resolution coverage across the full solar atmosphere, restricting vortex tracking to the photosphere and chromosphere [32-42]. Synoptic instruments lack the spatial resolution to detect vortices, leaving observations dependent on chance detections within the limited fields of view of high-resolution telescopes. This absence of continuous, high-resolution data hampers large-scale statistical studies. Furthermore, current techniques recover photospheric flows only under restrictive assumptions and perform poorly for the small, rapidly rotating, multidirectional motions characteristic of vortices [43-45], especially outside the photosphere. With no reliable method to recover chromospheric or transition region (TR) velocity fields, most observable vortex properties remain inaccessible.
- *Simulation limitations:* while vortices emerge naturally in high-resolution numerical models, most simulations still assume quiet-Sun conditions with simplified magnetic topology [46-49], leaving vortex dynamics in more complex environments largely unexplored (e.g., [10]). Many of the current state-of-the-art codes lack fully realistic radiative transfer see e.g. [50], which limits how well we can connect simulated vortices to observations and therefore constrains our ability to analyse or interpret their dynamics in real data. The behaviour of vortices in coronal holes (CHs) or active regions (ARs), or their evolution into the TR, low corona and beyond, is almost entirely unconstrained [10]. At present, high-fidelity synthetic observables can only be generated for a small number of photospheric and chromospheric line diagnostics, and usually for selected time frames [50]. As a result, simulations cannot yet be matched to observations across full spectral ranges, hindering robust validation.
- *Lack of understanding of vortices in ARs:* Perhaps the most significant gap is the absence of systematic studies of vortices in ARs, as the community has invested almost exclusively in the quiet Sun for vortical studies. Strong magnetic fields, steep gradients, shear flows, and complex current systems suggest that vortices in these environments may behave quite differently [51, 52] from their quiet-Sun counterparts with potentially significant roles in energy storage and release, current-sheet formation and reconnection triggering, jet or flare onset, and local heating or wave excitation [10, 13; 53-57). In other words, we are effectively blind to their role in the very regions where vortices matter the most. Over the coming decade, establishing the prevalence, spatial structure, lifetimes, and energetics of vortices in sunspots, plage, and emerging-flux regions will be transformative for heliophysics. In ARs, vortices can play key roles in magnetic restructuring and ultimately space-weather severity.
- *Solar wind origin understanding*: Another key question that has to be addressed is whether vortex-generated energy transport and magnetic twisting, particularly in ARs and CHs, can create the conditions [10, 21, 29] that produce both the bulk solar-wind outflow and the ubiquitous switchbacks detected by Parker Solar Probe. Pertinent to this is how multiscale vortical motion connects to large-scale heliospheric dynamics.

**We propose to test the paradigm-shifting idea that solar vortical plasma motions (individually and as a coherent, cross-scale network of their communities) form the dominant physical mechanism governing energy transfer, atmospheric heating, wave production and magnetic reconfiguration throughout the solar atmosphere.** In particular, we propose to export the advantages of the tunable Fabry-Perot Interferometers (FPI) instrumentation

into space technology to tackle the above challenges. **Our proposal requires coordinated progress in observations, theory, and simulation, areas where the UK is ideally placed to lead. The results will advance predictive understanding of solar variability and strengthen the UK's contribution to heliophysics, aligning closely with STFC's priorities for the coming decade**

Establishing how multiscale vortical motion connects to large-scale heliospheric dynamics will directly strengthen the UK's ability to forecast space-weather-relevant phenomena: an STFC and national priority. Vortices also provide a natural laboratory for key Theme 3 (Space Plasma Processes) questions, including wave behaviour in inhomogeneous plasmas (SP2), turbulence modelling (SP3), reconnection timescales and topology change (SP4), cross-scale coupling (SP6), and particle acceleration (SP7). Their multiscale, intermittent nature offers the framework needed to bridge gaps between MHD and kinetic processes identified as high-priority challenges for the UK community. Aligned with the STFC Solar System Research Roadmap (2022), Decadal Survey (NASA) and the UK's strategic ambition to lead in solar-variability and space-weather research, we highlight the following high-priority scientific questions for the coming decade. These address how small-scale plasma processes couple to global solar dynamics and heliospheric consequences, directly supporting Roadmap themes on multiscale coupling, magnetic-field evolution, turbulence, reconnection, and solar-wind generation.

*1. What fundamental plasma processes generate, sustain, and destabilise vortices, and how do these processes regulate the transfer of mass, momentum, and energy throughout the solar atmosphere?*
Vortices are a dominant [33-42], yet unresolved, component of solar surface dynamics. Understanding the physical mechanisms that generate and dissipate them is crucial for unlocking the physics that couple the photosphere, chromosphere, and corona. This is a frontier plasma-physics question with direct implications for turbulence, reconnection, and wave generation.

*2. How do vortices connect atmospheric layers through vertical coherence, and what physical factors determine their evolution, fragmentation, or dissipation as they traverse the solar atmosphere?*
Establishing the vertical evolution of vortices, from their sub-surface formation in the convection zone to their presence in the corona, addresses one of the most important unknowns in heliophysics: the pathways by which energy and magnetic stress are transported upward, which is also central to solving chromospheric and coronal heating [19].

*3. How do multiscale vortex systems restructure magnetic fields, concentrate currents, and drive the build-up and release of magnetic energy across quiet Sun, active regions, and coronal structures?*
This elevates the problem from local dynamics to Sun-wide energetics. Understanding how vortices shape magnetic topology [e.g.. 10, 21] at multiple scales is essential for connecting photospheric flows to coronal evolution, flare precursors, and long-term variability.

*4. Do interacting vortices create coherent structures, such as waveguides, flux channels, or turbulence networks, that fundamentally alter wave generation, mode conversion, and energy dissipation across the solar atmosphere?*
This question targets the heart of modern solar atmospheric physics: the interplay between waves, turbulence, and structured magnetic environments. Proving vortices form collective energy-transport systems, as suggested by their network dynamics established in hydrodynamics theory [58], would reshape our understanding of how heating, wave propagation, and dissipation occur.

*5. To what extent does vortex-driven magnetic restructuring act as a trigger for explosive events, such as jets, flares, and CMEs, and what is its role in generating and modulating the solar wind?*
This connects small-scale processes to major space-weather drivers. Determining whether vortices initiate reconnection, destabilise coronal structures, or inject mass and momentum into the solar wind addresses a long-term UK priority with major implications for space-weather forecasting.

## 2. Strategic Context

*Strategic Scientific Positioning:* The UK is exceptionally well-positioned to lead a coordinated international programme on vortex-driven solar physics. Building a focused national research framework around solar vortices would consolidate UK leadership in solar physics and allow the UK to shape future mission requirements, particularly the need for high-cadence, high-resolution spectropolarimetry and other magnetically sensitive diagnostics optimised for multiscale plasma flows. It would also stimulate innovation in inversion techniques, solar seismology, and data-driven modelling tailored to the fine-scale structuring that vortices produce. Such a programme would naturally position the UK to lead international collaborations centred on multiscale energy transport and cross-layer coupling, and to act as a scientific architect for the next generation of solar and heliophysics missions. With coordinated investment in observations, modelling, and instrumentation, the UK has the opportunity to establish itself as the global leader in understanding the small-scale plasma processes that govern the behaviour of the Sun and shape the conditions throughout the heliosphere. Investing in this research now positions the UK industry to benefit from next-generation models that incorporate the small-scale physics shaping the heliosphere. Solar storms, aurorae, and the solar cycle also attract strong public interest, making a national effort on solar vortices a natural platform for public engagement, STEM outreach, and education showcasing the UK's leadership in solar science.

## 3. Proposed approach

To harness the scientific potential of solar vortices and secure the UK's position as a leader in this emerging field, we propose a coordinated research strategy that synergetically integrates observations, numerical simulations, and theoretical advances. The overarching goal is twofold: first, to create a framework that enables existing and future observational infrastructure to capture vortex dynamics with the required accuracy; and second, to develop an analysis framework for numerical simulations that maximises the scientific return from high-resolution datasets.

A critical early priority is to unlock the true nature of velocity fields in the solar atmosphere. Current methodologies (FLCT, ML approach see e.g. [43, 45]) struggle to extract accurate horizontal vortical flows in the photosphere, particularly in the rotational and shear components required for reliable vortex detection ([42,45]). Even more critically, presently there is no robust method for measuring the chromospheric plane-of-sky velocity fields. This deficit renders key aspects of vortex dynamics effectively invisible, forcing us to rely on indirect simulation studies. Overcoming this severe limitation demands immediate, targeted investment in next-generation spectroscopic inversion techniques, advanced machine-learning flow-tracking, and magnetically sensitive multi-line diagnostics.

Despite ARs being the prime candidates for intensified vortex activity - characterised by strong magnetic gradients, shear flows and turbulence - systematic studies remain surprisingly scarce. A dedicated, coordinated observational-numerical effort is needed to quantify their true spatial scales, occurrence rates, and precise dynamical roles. This endeavour is essential to clarify how these small-scale motions feed into the broader magnetic evolution that underpins major solar activity. Furthermore, a significant, virtually unexplored frontier concerns vortex-related instabilities. Though barely explored under solar conditions, these instabilities have the profound potential to drive rapid reconfiguration of plasma and magnetic fields, potentially triggering jets or other impulsive phenomena. Progress hinges entirely on tightly integrated simulation and observational studies to confirm if such instabilities operate in the solar atmosphere and to identify their diagnostic signatures.

A paramount priority is to investigate the causal relationship between vortices and magnetic reconnection. The inherent topology and twisting motions of vortices are hypothesised to directly generate current sheets, drive shear flows across magnetic boundaries, or create favourable conditions for the onset of reconnection [4,21]. We can rigorously approach this challenge by integrating high-resolution multi-wavelength observations with data and physics-driven MHD simulations using HPC resources. This methodology requires selecting active regions that host large-scale vortices, accurately extrapolating their magnetic fields, and using the observed velocity fields as lower-boundary drivers in numerical models. This combined approach will enable us to precisely track the formation of current sheets, magnetic connectivity changes, and energy conversion associated with vortex evolution, helping to determine whether vortices directly trigger reconnection or instead enhance its efficiency. This is essential for quantifying their contribution to energy release throughout the solar atmosphere and evaluating their potential link to nanoflare-like phenomena.

Finally, the role of vortices as waveguides demands dedicated study [25,26]. Individual or even interacting vortices may form channels that guide or concentrate wave energy, with consequences for heating, transport, and atmospheric coupling. Key questions include identifying the types of waves supported by vortex structures in observations, determining how far such waves can propagate within a vortex, and understanding how wave-vortex interactions influence energy dissipation. Addressing these questions will require coordinated observations, modelling, and theoretical work to bridge the gap between small-scale plasma processes and global atmospheric dynamics. Taken together, this proposed approach establishes a comprehensive and forward-looking framework for advancing vortex science. It builds on UK strengths in high-resolution observations, spectropolarimetry, and plasma modelling while creating new pathways for discovery that align with national and international strategic priorities.

## 4. Proposed technical solution and required development

A coordinated technical pathway combining advances in instrumentation, data analysis, and numerical modelling is essential for enabling the UK to characterise multiscale vortices throughout the solar atmosphere. A key requirement is the development of methods that can reliably recover velocity fields in both the photosphere and chromosphere. Numerical modelling, addressed in a separate work package, plays a crucial role by providing high-resolution, realistic MHD simulations that resolve solar vortices. These simulations act as numerical experiments, revealing the observational signatures vortices produce and generating realistic synthetic data to guide instrument design, including optimal spectral lines, sensitivities, cadences, and sampling strategies. They also supply ground-truth flow fields for validating velocity-reconstruction techniques and assessing their performance limits before hardware development begins.

Our long-term mission is to establish a unified observational and numerical framework to ensure that future capabilities are tailored to the physics of vortical structures. Building these analysis frameworks, however, will only succeed if matched by next-generation observational capability. The main limitation of current facilities is not spatial resolution, but the lack of co-temporal, high-cadence, multi-line spectropolarimetric diagnostics needed to recover velocities, magnetic fields, and thermodynamic parameters. This requires high-cadence, narrow-band imaging spectroscopy across multiple lines on sub-second timescales, together with spectropolarimetric sensitivity in key chromospheric lines such as Na D1, K I, Ca II 8542, He I 10830, and Mg II h&k.

Finally, investigating vortex dynamics demands robust multi-height diagnostics. Currently, our ability to track vortices into the transition region or low corona is limited because existing high-resolution observations (with cadences of seconds and resolution of a few kilometres) probe mostly the low atmosphere. Future capability must therefore prioritise spectral lines with narrow, well-defined formation heights, since these allow us to sample distinct atmospheric layers cleanly and follow vortex evolution without ambiguity. Additional priorities include continuous, high-cadence, high-resolution coverage from the deep photosphere to the low corona, and simultaneous multi-line acquisition to avoid losing rapidly evolving signatures to sequential scanning. Achieving this requires optical systems that isolate spectral lines with high purity and switch rapidly between them, including advances in filtergraph design, fast-tuning etalons (tunable Fabry Pérot filters), and next-generation magneto-optical filters or integral-field spectropolarimetry. Developing such instrumentation would not only enable vortex studies but also provide critical data for probing other small-scale and eruptive activity, including nanoflares, EUV brightenings, campfires, flare ribbon substructure, and jet-like phenomena such as spicules, mottles, and jetlets. A balloon-borne demonstrator mission offers the most efficient path to maturing these technologies, providing a low-distortion environment to validate instrument architectures before space deployment. To maximise scientific return and prepare for future missions, the UK should invest in data pipelines and coordinated observing frameworks that integrate with DKIST, Solar Orbiter, MUSE, SOLAR-C EUVST, and SST

## 5. UK leadership and capability

The UK has an exceptionally strong and diverse solar-physics community, with expertise distributed across a wide network of universities and national laboratories in England, Scotland, Wales, and Northern Ireland. Research groups at Imperial College London, UCL/MSSL, the University of Aberystwyth, Birmingham, Cambridge, Dundee, Durham, Exeter, Glasgow, Manchester, Northumbria, Sheffield, St Andrews, Queen's University Belfast, and the Armagh Observatory, among others, constitute one of the most comprehensive national ecosystems for solar, plasma, and space-weather science worldwide. This distributed research base is complemented by RAL Space, which provides world-class engineering, calibration, and mission-delivery capability, anchoring the UK's technical strengths in space instrumentation.

The UK has played a pioneering and sustained role in advancing the study of solar vortices, contributing internationally recognised work in high-resolution imaging, spectropolarimetry, magnetoconvection theory, plasma turbulence, and vortex dynamics. UK researchers were among the first to identify small-scale rotational flows in the photosphere and chromosphere, and they continue to lead efforts to understand their energetics, wave generation, cross-layer coupling, and links to magnetic restructuring. Theoretical advances in MHD turbulence, vortex formation, and instability from groups in Sheffield, St Andrews, Dundee, Glasgow, Northumbria, and Exeter have shaped the global agenda on multiscale solar dynamics, establishing the UK as a conceptual leader in this emerging field. This leadership is further demonstrated by the UK's foundational role in establishing the Solar Physics International Network for Swirls (SPINS), which builds on pioneering UK-led collaborations with world-leading institutes in solar-vortex research. In parallel, the UK has developed a world-leading position in space-weather science and operations, providing a strong bridge between fundamental research and real-world application. The Met Office Space Weather Operations Centre (MOSWOC) remains Europe's foremost national forecast centre, underpinned by extensive university expertise. UK leadership in ESA's Solar Orbiter mission, major contributions to DKIST instrumentation, and prominent involvement in forthcoming missions such as MUSE and SOLAR-C EUVST ensure privileged access to the highest-quality solar datasets of the coming decade and place the UK at the forefront of heliophysical discovery.

Instrumentation is a core UK strength. UK teams have delivered major hardware for Solar Orbiter and key detector and camera systems for DKIST, supported by RAL Space and industry partners such as Teledyne e2v and Leonardo, who provide world-class manufacturing and calibration expertise. Growing national capability in adaptive optics, high-speed polarimetry, and lightweight optical designs positions the UK to drive next-generation solar instrumentation, including high-cadence multi-line spectropolarimeters and chromospheric vector-magnetic diagnostics, needed to resolve multiscale vortex dynamics. Combined with strong national expertise in observations, modelling, theory, and space-weather operations, this instrumentation base gives the UK a uniquely integrated platform for future leadership. A coordinated national programme would enable the UK to shape upcoming missions, lead international campaigns, and deliver transformative advances in understanding the small-scale plasma processes that govern solar variability and heliospheric conditions.

## 6. Partnership Opportunities

A UK-led programme on vortex-driven solar physics would naturally integrate with a broad and diverse set of international partners. Close collaboration with the EST, ESA Solar Orbiter community and the extensive DKIST international networks would ensure access to the highest-resolution observations currently available, while engagement with the US-led MUSE and Japan-led SOLAR-C EUVST teams would align UK efforts with the major next-generation space missions now in development. Our current collaboration with the Swedish 1-m Solar Telescope and Sunrise III teams, involving the provision of data, analyses, and high-resolution diagnostics pertinent to vortical studies, provides a strong foundation for expanding synergies in instrumentation development where these partners excel. Further strategic depth comes from collaboration with institutions such as the Instituto de Astrofísica de Canarias (IAC, Spain), Indian Institute of Astrophysics (IIA, India), National Institute for Space Research (INPE, Brazil), National Observatory of Athens (NOA, Greece), Indian Institute of Science Education and Research, Pune (IISER, India), Institute for Space-Earth Environmental Research (ISEE, Japan), and the Italian National Institute for Astrophysics (INAF), each of which brings complementary strengths in instrumentation, modelling, or coordinated observing campaigns. Engagement with theoretical vortex-dynamics communities, spanning terrestrial, planetary, and astrophysical domains, would provide a rigorous cross-disciplinary framework for interpreting small-scale plasma processes. Together, these partnerships enable coordinated observations, shared modelling benchmarks, and joint development of future mission proposals, ensuring UK leadership within a strong and internationally visible collaborative ecosystem.

## 7. Suggested Mission Class

We aim to develop a first-of-its-kind, multi-band, space-qualified solar instrument that employs four tunable FPI to deliver small FOV, high-resolution (with diffraction-limited resolution of 150km in the infra-red, up to 30km in the vacuum-ultraviolet), spectral and spatial data. These will provide pseudo-simultaneous spectral coverage from the photosphere to low corona via the Infra-red (800-900 nm, covering Ca II IR triplet), optical (600-700 nm, covering Hα), near-UV (310–410 nm, covering Ca II H&K and Hε) and vacuum-UV (100–200 nm, covering Mg II h&k and Si IV). This capability directly overcomes the key observational barriers identified in Section 1. In addition, it provides the necessary data to study vortex evolution, wave guidance, reconnection drivers and cross-scale coupling.

We note that FPI technology is well established, and it has already been used in space. For example, the Polarimetric and Helioseismic Imager PHI on Solar Orbiter employs a non-tunable FPI. However, the integration of four tunable FPIs covering IR, optical, near-UV and VUV bands has never been realised in space, and thus constitutes a significant technological leap and scientific innovation. Such a multi-FPI system, although driven by the necessity to better understand vortical structures, is absolutely novel for space missions and also has immense scientific merit for studies involving instabilities and other small-scale phenomena and eruptive activity, including nanoflares, EUV brightenings, campfires, flare ribbon substructure, and jet-like phenomena (spicules, mottles, jetlets). This instrument could also be included in a multiple-instruments suite on a larger future space mission, designed to co-observe different-sized but co-aligned fields-of-view in an attempt to link how small-scale dynamics propagate or evolve to larger scales in the corona and beyond, such as switchbacks of the magnetic field recently observed in the solar wind mainly above open magnetic field line configurations.

The first mission stage is a high-altitude balloon platform to test the proposed FPI instrumentation under near-space conditions, including the coatings, detectors, control, durability and calibration. This would provide a low-cost, recoverable, upgradeable method to rigorously validate the later-stage mission concepts. We would validate the polarimetric sensitivity, thermal-control hardware, pointing and onboard data handling. It would also provide unique science data from the spectral-imaging datacubes provided by the Ultraviolet FPIs that could be used for scientific analyses in conjunction with coordinated space- and ground-based campaigns. The second mission stage would be to fly the developed instruments using a 1-m Balloon telescope to further validate the instrumental resolution, cadence, onboard data handling, polarisation-modulation schemes and coordinated multi-line acquisition strategies to achieve co-spatial measurements of photospheric magnetic fields and spectroscopic diagnostics across the lower solar atmosphere. The current state-of-the-art solar balloon instrumentation, SUNRISE-III, already demonstrates what is technically feasible: diffraction-limited resolution down to around 60 km at 309 nm and around 154 km at 855 nm [59], combined with cadences of only a few seconds for narrow-field scans and an exceptionally stable pointing performance of better than 0.05" sustained over hours. These are likely the practical upper limit of performance by balloon platforms using present-day optics and stabilisation systems. It is still not defined whether such performance is achievable for a broader spectral range with more demanding polarimetric requirements. It further provides an opportunity to quantify the computing resources required for multi-instrument co-registration. For example, SUNRISE-III produced over 200 TB of raw data over 6.5 days, illustrating that onboard systems must efficiently manage and compress to ensure the data can be stored and later fully processed on the ground. A balloon test flight, therefore, provides a realistic environment in which to evaluate whether future missions can reliably reach or exceed this benchmark. Our grand ambition is the future deployment of a high-resolution 1-m space-based telescope with such instrumentation, in collaboration with other national and international space agencies, that will provide an unprecedented spatial- and spectral-resolution view of the lower solar atmosphere from the photosphere to lower corona.

**Co-authors and signatories (in alphabetical surname order):**

| | |
|---|---|
| Mashael Aldhafeeri | University of Sheffield |
| Gert Botha | Northumbria University |
| Ioannis Dakanalis | National Observatory of Athens, Greece |
| Abhirup Datta | Indian Institute of Technology Indore |
| Robertus Erdelyi | University of Sheffield, UK |
| Mykola Gordovskyy | University of Hertfordshire |
| Andrew Hillier | University of Exeter |
| David Jess | Queen's University Belfast |
| Elena Khomenko | Instituto de Astrofisica de Canarias (IAC, Spain) |
| Matías Koll Pistarini | Instituto de Astrofisica de Canarias (IAC, Spain) |
| Anshu Kumari | Physical Research Laboratory, India |
| Hidetaka Kuniyoshi | Northumbria University, UK |
| Murali G. Meena | Oak Ridge National Laboratory, USA |
| Sargam Mulay | University of Glasgow, UK |
| Mohamed Nedal | Dublin Institute for Advanced Studies (DIAS), Ireland |
| Divya Paliwal | Physical Research Laboratory, India |
| Thomas Rees-Crockford | Northumbria University, UK |
| Luiz Schiavo | Northumbria University, UK |
| Azaymi Siu | Instituto de Astrofisica de Andalucia (IAA-CSIC), Spain |
| Samuel Skirvin | Northumbria University, UK |
| Shivdev Turkay | Northumbria University |
| Nitin Yadav | Indian Institute of Technology, Delhi, India |
| Jinge Zhang | Paris Observatory - PSL, France |
| Sergii Zharkov | Hull University, UK |
| Valery Nakariakov | Warwick University, UK |